# Wavelet Based Load Models from AMI Data

Shiyin Zhong, *Student Member IEEE*, Robert Broadwater *Senior Member, IEEE*, Steve Steffel, *Senior Member.*

*Abstract*—A major challenge of using AMI data in power system analysis is the large size of the data sets. For rapid analysis that addresses historical behavior of systems consisting of a few hundred feeders, all of the AMI load data can be loaded into memory and used in a power flow analysis. However, if a system contains thousands of feeders then the handling of the AMI data in the analysis becomes more challenging.

The work here seeks to demonstrate that the information contained in large AMI data sets can be compressed into accurate load models using wavelets. Two types of wavelet based load models are considered, the multi-resolution wavelet load model for each individual customer and the classified wavelet load model for customers that share similar load patterns. The multi-resolution wavelet load model compresses the data, and the classified wavelet load model further compresses the data. The method of grouping customers into classes using the wavelet based classification technique is illustrated.

## I. Introduction

Power flow analysis, as well as many other power system analysis applications, can benefit from load data collected from Advanced Metering Infrastructure (AMI). Planning, forecasting, automated customer type identification and classification, real-time analysis, and even real-time control can benefit from information derived from AMI load data.

Some previous efforts in using customer load measurements in power system analysis have used load research statistics to generate 8760 hourly statistical load models for classes of customers [1]. The AMI implementation gives a utility the ability to collect each individual customer's detailed load profile. With such detailed load data the accuracy of load models can be improved [2].

A utility normally collects load data from each customer through AMI every 15 to 60 minutes. Thus, each customer can have 8,760 to 35,040 rows of load data annually. For a small circuit with 100 customers, an annual AMI load data table in a database can contain from 876,000 hourly load data rows to 3,540,000 million rows.

The work here focuses on deriving load models that can be placed into computer memory and used in power flow time-series analysis in large scale models. The sheer size of the AMI load data sets makes the direct AMI data integration somewhat impractical for large scale systems. For example, using double precision numbers a utility with 1 million customers would require 1,000,000 × 8 bytes × 8,760 hours/year ≈ 65G bytes of RAM to store the annual hourly AMI load data in memory.

The development of wavelet transformations began with Alfréd Haar's work in the early 20$^{th}$ century [3]. The wavelet related research accelerated after the ground breaking works from Ingrid Daubechies [3] and Stéphane Mallat [4] in the 1980s. There are many applications of wavelets, including digital signal processing and image processing [5].

Two wavelet based load models are used in this paper. The first load model is the multi-resolution wavelet load model, which will often be referred to here as the wavelet load model. It uses the Discrete Wavelet Transformation (DWT) [4] to transform the original load profile from the time domain to the wavelet domain. With the multi-resolution wavelet load model each individual customer's AMI data is compressed, and load models are maintained for each individual customer.

The second load model considered here is the classified wavelet load model. With the classified wavelet model a single load model is used for many customers that exhibit similar load behavior. With wavelet based load models, the determination of which class a given customer should be assigned to can be automated.

Conventionally the classification of a load profile is based on either the load profile's time domain [6] or frequency domain descriptors [7]. Time domain parameters used to describe load profile characteristics include Base Load, Peak Load, Rise Time, Fall Time, and High-Load Duration. After a set of variables is established to describe load profiles, various models, such as linear discriminant analysis, nearest neighbor classification, k-means, fuzzy-statistics, neural networks, and support vector machines are used to classify load profiles [8-14].

A major challenge in classifying load profiles is that individual customer' AMI load data may contain extreme data changes, such as the load rapidly going to zero.

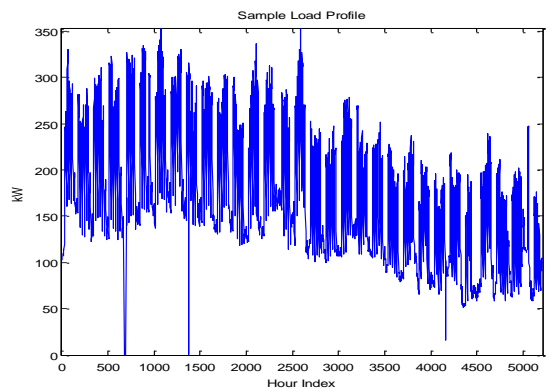

**Figure 1: Sample AMI Load Profile Data**

Sometimes these extreme changes are errors, but not always. Figure 1 illustrates representative AMI hourly load samples for a commercial customer. The aforementioned classification methods, where relatively small variable sets are employed, have trouble in accurately capturing the pattern characteristics of Figure 1.

In order to better visualize the load profiles' pattern characteristics, a 2 dimensional, or 2D, wavelet load profile representation is introduced in this paper. It will be illustrated that the significant load patterns in an individual customer's AMI load data can be modeled by the wavelet components from the load profile's lower 2D DWT transformation. Furthermore, the 2D DWT transformation can be used in an unsupervised clustering process to identify load classes. The classified wavelet load models are derived from the load classes identified by the 2D DWT based clustering algorithm.

In Section II both 1D and 2D multi-resolution wavelet decompositions are discussed. In Section III the procedure of compressing AMI load data using 1D multi-resolution wavelet decomposition coefficients is introduced. After that, the k-mean clustering algorithm using 2D wavelet decomposition coefficients is presented. In Section IV three sets of analysis results for a real-world circuit are compared. The comparisons utilize time series power flow analysis where the base case runs over the original AMI load measurements, and the other cases use the wavelet load model and the classified wavelet load model, respectively. Conclusions are presented in Section V.

## II. DISCRETE WAVELET TRANSFORMATION

In the DWT process, a time-domain discrete signal $S(n)$ can be decomposed into a set of Approximation Coefficients $A_{j_0,k}$ (equation 1) and Detail Coefficients $D_{j,k}$ (equation 2) with a predetermined discrete scaling function $\varphi_{j,k}(n)$ (equation 3) and wavelet function $\psi_{j,k}(n)$ (equation 4). The $S(n)$ can be reconstructed using its $A_{j_0,k}$ and $D_{j,k}$ in the inverse DWT (IDWT) process (equation 5).

$$A_{j_0,k} = M^{-\frac{1}{2}j} \sum_n S(n)\varphi_{j_0,k}(n) \qquad (1)$$

$$D_{j,k} = M^{-\frac{1}{2}j} \sum_n S(n)\psi_{j,k}(n) \qquad (2)$$

$$\varphi_{j,k}(n) = 2^{j/2}\varphi(2^j n - k) \qquad (3)$$

$$\psi_{j,k}(n) = 2^{j/2}\psi(2^j n - k) \qquad (4)$$

$$S(n) = M^{-\frac{1}{2}j} \sum_k A_{j_0,k}\varphi_{j_0,k}(n) + \sum_{j=0}^{\infty}\sum_k D_{j,k}\psi_{j,k}(n) \qquad (5)$$

where n=0,1,...,M-1 and M=$2^J$

decomposing level $j = 1,...,J-1$, $j_0 = 0$

coefficient index $k = 0,1,...,2^{j-1}$

$\varphi_{j_0,k}(n)$, $\psi_{j,k}(n)$ are the scale and wavelet function

### A. Multi-Resolution DWT

Mallat introduced an efficient multi-resolution DWT/Inverse DWT (IDWT) algorithm in 1989 [4] which made the DWT/IDWT implementation practical by taking advantage of the family of orthogonal, compact support wavelets introduced by Daubechies [3].

Mallat's pyramid DWT/IDWT algorithm uses a set of discrete Quadrature Mirror Filters (QMF) ($h_\psi, h_\psi, h_\varphi, h_\varphi$) to decompose or reconstruct a discrete signal. The coefficients of this set of filters are pre-determined. The simplest Daubechies (DB) wavelet filters, DB1 (Haar wavelet), is used in the AMI load data DWT here. The DB1 scale and wavelet discrete filter coefficients are shown in

| | |
|---|---|
| Wavelet | $h_\psi = \left[-1/\sqrt{2}, 1/\sqrt{2}\right]$ |
| | $h_\psi = \left[1/\sqrt{2}, -1/\sqrt{2}\right]$ |
| Scale | $h_\varphi = \left[1/\sqrt{2}, 1/\sqrt{2}\right]$ |
| | $h_\varphi = \left[1/\sqrt{2}, 1/\sqrt{2}\right]$ |

**Table 1: DB1 QMF Coefficients**

table 1.

In the work that follows $S[n]$ will represent the AMI data set for a single customer. Figure 2 illustrates a 3-level DWT process. The discrete signal $S[n]$ convolutes with the low band-pass filter $h_\varphi$ (covers 0 to $f_n/2$ frequency band as in figure 3) which generates $n$ number of values. Half of this set of numbers is redundant. Therefore, it can be down-sampled by 2 to obtain the Approximation Coefficients $A_j$. The down-sampling (decimation) process involves removing every other coefficient from the $A_j$ approximation coefficients.

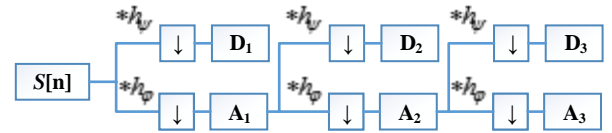

**Figure 2: Multi-resolution DWT**

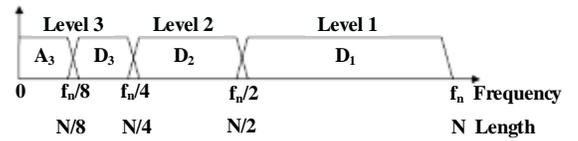

**Figure 3: Multi-resolution DWT Frequency Bands**

$S[n]$ also convolutes with the high band-pass filter $h_\psi$ (covers $f_n/2$ to $f_n$ frequency band) and is then down-sampled by 2 to calculate the Detail Coefficients $D_j$. The approximation coefficients $A_j$ are then passed to the next level to repeat the same transformation process to generate the $j+1$ level $A_j$ and $D_j$ as specified in equations 6 and 7.

$$D_{j+1}[n] = \sum_{k=0}^{N} h_\psi[k]A_j[2n-k] \quad (6)$$

$$A_{j+1}[n] = \sum_{k=0}^{N} h_\varphi[k]A_j[2n-k] \quad (7)$$

where N is the length of Finite Impulse Response Filters $h_\varphi, h_\psi$

Figure 4 illustrates a 3-level IDWT process which reverses the DWT process to reconstruct the $S[n]$ using $A_j$ and $D_j$.

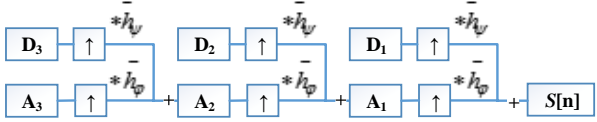

**Figure 4: Multi-Resolution IDWT**

During the IDWT process, the coefficients $A_j$ and $D_j$ at level j are up-sampled by 2 and the up-sampled coefficients convolute with the mirror discrete filters $h_\varphi$ and $h_\psi$ respectively. The two convolution products are added together to generate coefficients $A_{j-1}$ at level j-1 as in equation 8. These steps are repeated until the $S(n)$ is fully reconstructed.

$$A_{j-1}[n] = \sum_{k=0}^{N} h_\varphi[k]A_j[2n-k] + \sum_{k=0}^{N} h_\psi[k]D_j[2n-k] \quad (8)$$

where N is the numboer of coefficients in $h_\varphi, h_\psi$

The DWT/IDWT algorithm described by equations (6-8) has a linear computational complexity $O(n)$ that needs $n$ operations to decompose a discrete load profile (n number of measurements) into the wavelet domain.

Figure 5 presents a 3 level wavelet representation for a sample AMI load profile (with 5,208 measurements). The plots in the left column are the synthesized load profiles $S_{A_j}(n)$, which are reconstructed (equation 9-10) by using only the approximation coefficients $A_j$ at level j. The $D_j$ coefficients are not included in the synthesizing process.

$$S_{A_j}[n] = \sum_{k=0}^{N} h_\varphi[k]A_1[2n-k] \quad (9)$$

$$A_{j-1}[n] = \sum_{k=0}^{N} h_\varphi[k]A_j[2n-k] \quad (10)$$

where N is the numboer of coefficients in $h_\varphi$

$j \in [2, J]$ where $J$ is the largest decomposition level.

This process is illustrated in figure 6 which shows that detail coefficients are eliminated from the simplified IDWT. These synthesized load profiles cover the lower frequency sub-band of the original AMI load profile.

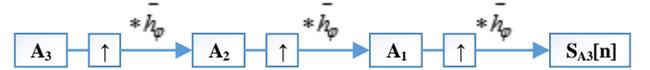

**Figure 6: Synthesized load profile $S_{A_3}[n]$**

The second column plots are the synthesized $S_{D_j}(n)$, which is reconstructed (figure 7) with only the detail coefficient $D_j$ at level j. The $A_j$ coefficients are not included in the synthesized process. These synthesized load profiles represent the higher frequency information in the original AMI load profile

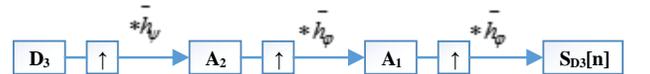

**Figure 7: Synthesized load profile $S_{D_3}[n]$**

The synthesized level j approximation load profile $S_{A_j}[n]$ needs twice as many coefficients as the level $j+1$ synthesized load profile $S_{A_{j+1}}[n]$. In figure 5, the $S_{A_1}[n]$ load profile is modeled by $5,208 / 2^1 = 2,604$ coefficients, the $S_{A_2}[n]$ load profile is modeled by $5,208 / 2^2 = 1,302$ coefficients, and the $S_{A_3}[n]$ load profile is modeled by $5,208 / 2^3 = 651$ coefficients, which is 1/8 the size of the original load profile (figure 1).

From visual inspection the level 3 synthesized load profile $S_{A_3}[n]$ is very similar to the original load profile presented in figure 1. In section 4 it will be illustrated that the majority of the energy content of the signal can be captured by $S_{A_3}[n]$.

Wavelet coefficients can be used in frequency-time localization analysis, such as the detection of large sudden changes in a load profile. In figure 8 a sample load profile and its 2 level DWT detail coefficients are presented. There are two sudden and significant load drops in the load profile: the first one takes place from the 51$^{st}$ to the 52$^{nd}$ hour, and the second one occurs at the 73$^{rd}$ hour. The 73$^{rd}$ hour drop has a higher frequency since its period is shorter. Once the load profile is transformed into the wavelet domain with 2-levels of resolution, the two load drops can be detected by analyzing the two detail coefficients sets ($D_1$ and $D_2$ plot in figure 8). The $D_1$ coefficient from the level 1 DWT covers the higher half of the frequency band ($f_n/2$ to $f_n$). The higher frequency $73^{rd}$ hour load drop can be observed when the $D_1$ magnitudes suddenly drop below -100 at the $74^{th}$ hour. The lower frequency $51^{st}$ to $52^{nd}$ hour load drop can be detected by analyzing the $D_2$ coefficients, which covers the $f_n/4$ to $f_n/2$ frequency band. In this case the $D_2$ coefficient spikes to over 100 at the $51^{st}$ hour when the load drop occurs. In this example the approximation coefficients are left out because the sudden load changes are high frequency signals that should not be visible in the lower frequency band where the approximation coefficients reside.

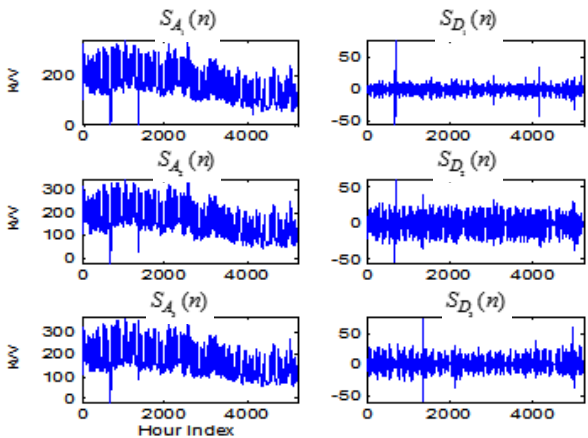

**Figure 5: Three Level DWT using Haar wavelet**

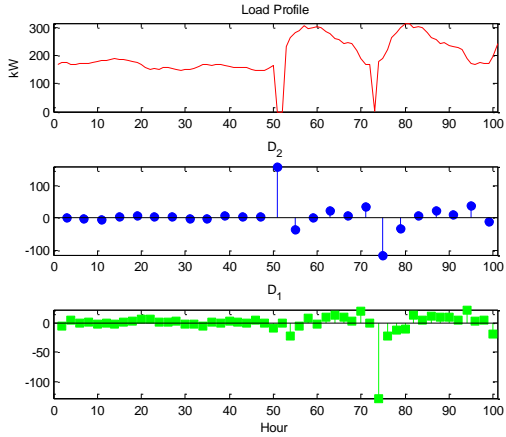

**Figure 8: Load Profile Detail Coefficient from 2-Level DWT**

### B. 2D Wavelet Transformation

The 2D wavelet decomposition process is similar to the 1D DWT decomposition as the signal goes through a series of discrete wavelet filters $h_\psi, h_\varphi$. The 2D discrete wavelet decomposition first decomposes the signal in columns, followed by a decomposition into rows, as illustrated in Figure 9. At each level, the 2D wavelet transformation will generate one approximation coefficient $W_\varphi(j)$ and three different detail coefficient sets: the horizontal $w_\psi^h(j)$, the vertical $w_\psi^v(j)$ and the diagonal coefficients $w_\psi^d(j)$.

The reconstruction process reverses the decomposition process which reconstructs the data by performing the IDWT along the row direction, followed by performing the IDWT along the column direction.

### III. LOAD DATA MODELING

The real-world circuit used in this paper serves 323 customers whose hourly AMI load measurements were recorded from May 2013 to Dec 2013. These customers represent a diverse group of load patterns. For example, there are customers with the double-peak residential load type, businesses with the typical 9am to 5pm commercial load type, loads with on-off type behavior (similar to street light load patterns), and others. Identifying the various load patterns for the circuit will be considered shortly.

In the analysis AMI measurements from 2013-05-27 00:00 to 2013-12-29 23:00 are used, representing a 31 week time window. Each customer had 5208 hourly kWHr measurements, for a total of 1,682,184 rows of AMI measurements for the 323 customers served by the circuit.

### A. Multi-resolution Wavelet Load Model

For time-series analysis involving AMI data, SCADA data, and others, it is desirable to load the data into computer memory for the most rapid analysis. However, for models consisting of thousands of circuits with similar AMI coverage it is not be practical to load all of the AMI data directly into computer memory. The goal of a compressed AMI load data model is to construct a load model that accurately approximates the original load profile's energy and significant pattern characteristics with a much smaller data set.

Typical load profile patterns tend to be relatively slow changing, indicating that most of the load energy is located in the lower frequency band, which can be represented by the approximation coefficients $A_j$ introduced earlier. The sudden, abnormal changes in a load profile have much higher frequencies which are modeled by the $D_j$ detail coefficients. The high frequency behavior modeled by the detail coefficients generally represents a small portion of the load profile's energy. From figure 3 it may be noted that the $S_{D_j}(n)$ load profiles have much smaller magnitudes than the corresponding $S_{A_j}(n)$ load profiles.

In this paper the $S_{A_3}(n)$ load profile is used as the load model. This model has a coefficient set that is 1/8 of the original AMI data size. The model retains most of the energy residing in the original load profile and captures much of the original load profile's characteristics. However, it should be noted that it is possible to model the AMI load data at either higher (more coefficients) or lower (less coefficients) resolutions in the wavelet domain. But in this paper characteristics of using $S_{A_3}(n)$ will be investigated.

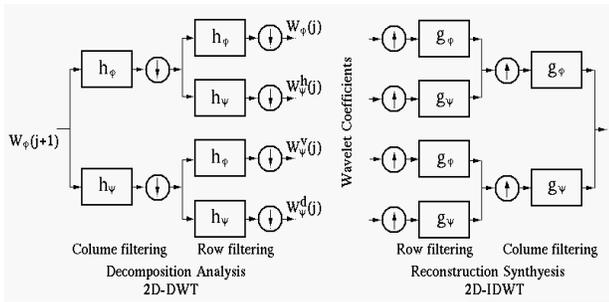

**Figure 9: 2D DWT**

In order to evaluate the performance of the $S_{A_3}(n)$ load model, three parameters are used which are: the synthesized load profile's total energy content compared to the energy content of the original AMI data; the distribution of the absolute value of the hourly error between the synthesized load profile and the original AMI load data; and the percentage error between the synthesized load profile and the original AMI load data. First, the variable SE defined in equation 11 is used to evaluate the synthesized load profile's total energy as a percentage of the original load profile's total energy.

$$SE = \left[\sum_{n=1}^{N} S_{A_1}(n)^2 \bigg/ \sum_{n=1}^{N} S(n)^2\right] \times 100\% \quad (11)$$

Table 2 summarizes the values of SE for all 323 load profiles synthesized from $S_{A_3}(n)$. In the worst case, three $S_{A_3}(n)$ synthesized load profiles captures less than 80% of the total energy of their original load profiles. Seventeen of the synthesized load profiles have a total energy that falls between 80 to 90% of the total energy of the original load profiles. The majority of the synthesized load profiles capture 90% or more of the original AMI load data energy.

| Number of $S_{A_3}(n)$ Load Profiles | SE |
|---|---|
| 3 load profiles (0.93%) | Less than 80.000% |
| 17 load profiles (5.26%) | 80.000% to 90.000% |
| 303 load profiles (93.81%) | 90.000% to 99.998% |

Table 2: Percentage of total energy modeled by load profiles synthesized from $S_{A_3}(n)$

AMI load data contains sharp load changes that are similar to the spikes previously discussed for figure 8, and the energy contained in these spikes is modeled with the detail coefficients and not the approximation coefficients of $S_{A_3}(n)$. The energy contained in these spikes is neglected when just $S_{A_3}(n)$ is used to synthesize the load profile. Thus, the energy errors shown in Table 2 are an indication of the variability of the loads, where three of the loads (i.e., those with an energy match less than 80%) have a much higher variability than the other 320 loads.

The second parameter used to evaluate the performance of the $S_{A_3}(n)$ load model is the hourly error, $E(n)$, which is defined as the difference between the synthesized load profile value and the original load profile value at each time point n, as given by

$$E(n) = \left|S_{A_3}(n) - S(n)\right| \quad (12)$$

Figure 10 presents a histogram for $E(n)$. A normal curve fit to the $E(n)$ distribution is shown as a red line in Figure 10. From Figure 10, it may be seen that the distribution of $E(n)$ is approximately normal with a mean that is centered close to zero and a small variance. This indicates that the error between the wavelet based load model and the original load data is small and predictable.

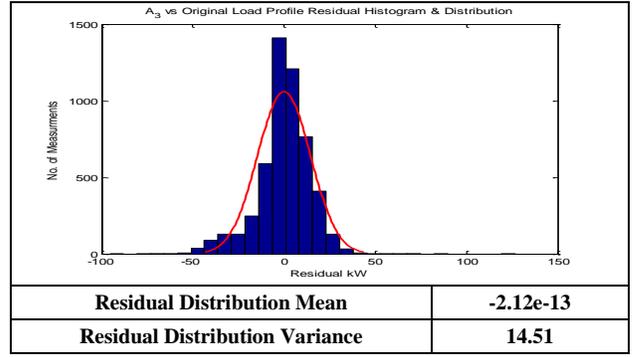

| Residual Distribution Mean | -2.12e-13 |
|---|---|
| Residual Distribution Variance | 14.51 |

Figure 10: Error of Synthesized Load Profile

The third parameter used to evaluate the performance of the $S_{A_3}(n)$ load model is the hourly percentage error, $PE(n)$, defined as

$$PE(n) = \left|\frac{S_{A_3}(n) - S(n)}{S(n)}\right| \times 100\% \quad (13)$$

Figure 11 presents the histogram for $PE(n)$. The result shows that 74% of the $S_{A_3}(n)$ hourly loads are within 10% of the original AMI load measurements.

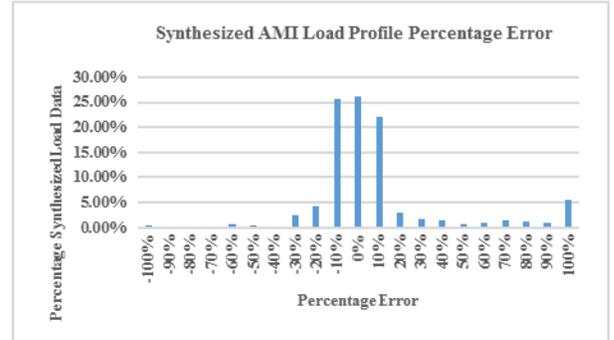

Figure 11: Histogram of Percentage Synthesized $S_{A_3}(n)$ Hourly Load Model Data with Percentage Error Less Than Given x-axis Value

### B. Classified Wavelet Load Model

With the classified wavelet load model individual customer load profiles are placed into groups, where the customers in each group have similar load patterns. The coefficients from each customer's 2D DWT are used to classify customers into groups.

In performing the grouping each of the 323 load profiles are normalized by its annual peak and converted into a matrix. In the matrix each row represents a week of hourly load measurements, from Monday at the 00 hour to Sunday at the 23rd hour. Thus, for the original AMI load data each row has 168 hourly readings. With 31 weeks of AMI data, the hourly AMI load profile is represented as a matrix with 31 columns and 168 rows. A similar matrix is built for the $S_{A_3}(n)$ synthesized load model.

The matrix representation of the load data can be visualized as a set of color images as illustrated in figure

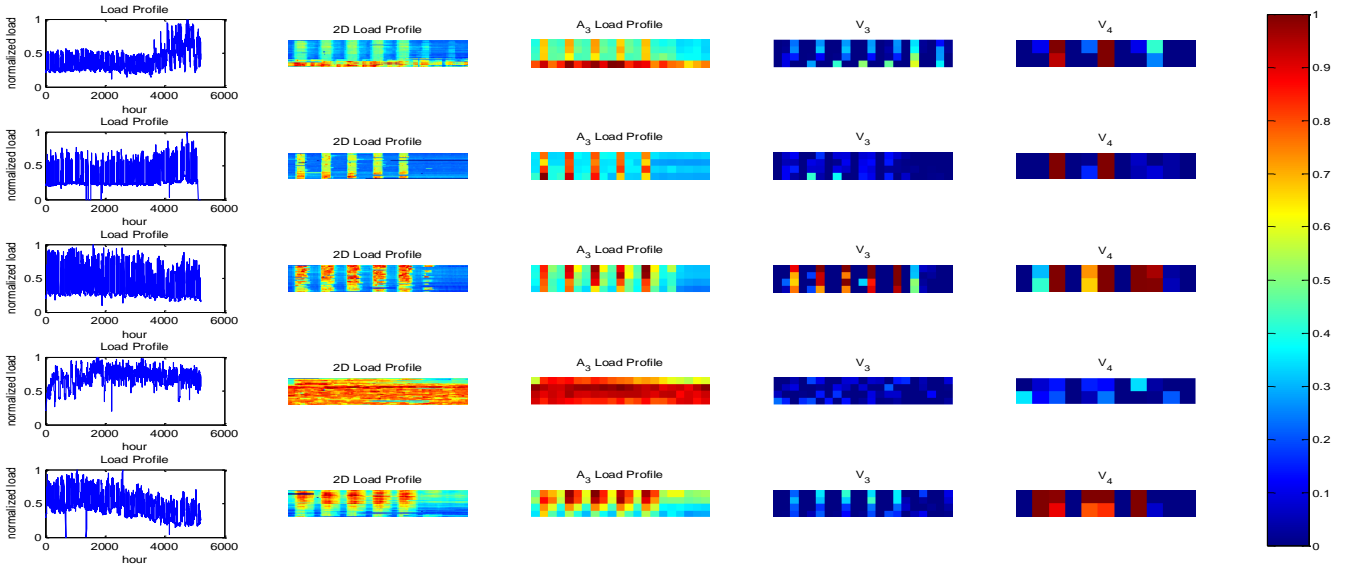

**Figure 12: Normalized load data color pattern visualizations where color map is shown on far right: First column is original AMI load data; Second column is color map visualization of original AMI data; Third column is color map visualization of synthesized $S_{A_3}(n)$ load model; Fourth and Fifth columns are color maps of Vertical Coefficients of the 2D DWT at levels 3 and 4, respectively**

12. All visual images share the color map shown on the right of figure 12. The plots in the first column of figure 8 represent the original normalized AMI load profiles. The second column in figure 12 presents a color map of the original AMI load data. The third column shows a color map for the $S_{A_3}(n)$ synthesized load model, which is a lower resolution of the second column's image. The fourth and fifth columns provide a visualization of the Vertical Coefficients of the 2D DWT at levels 3 and 4, respectively.

Some research proposes using Rise Time and Fall Time [2] to describe load patterns in the time domain. These sudden change in load profiles can be visually detected in the color maps of Figure 12 and also can be detected using the 2D DWT coefficients.

The fourth row in figure 12 illustrates a relatively consistent load profile without significant hourly load variations. Its level 3 vertical coefficient, $V_3$, shows a very consistent pattern without significant color variation. The other four load profiles in figure 7 all show daily patterns.

The daily pattern differences in load profiles may be detected by using the vertical coefficients shown in columns 4 and 5 of figure 12. The level 4 vertical coefficients, $V_4$, can be viewed as a "zoom-out" of $V_3$, providing a lower resolution, but still capturing the significant patterns of the original load profile. Instead of using 5,208 hourly load data measurements to classify customers, the more compact form of the DWT load pattern representations are used in the classification process. The size of the $V_3$ matrix size is 4 by 21 and the size of the $V_4$ matrix is 2 by 11.

A K-mean clustering algorithm [17] is used in the classification. The *K*-mean algorithm groups individual customer load profiles into one of k load classes, where the number of desired load class k is an input to the algorithm. The algorithm minimizes the mean of the Euclidean distance $D_{V_i, C_k}$ between the load profile Vertical Coefficients $V_i$ and the load class centroid $C_k$ (equation 14) for each class, as indicated by

$$D_{V_i, C_k} = \sqrt{\sum_{n=1}^{N}(V_{i,n} - C_{k,n})^2} \qquad (14)$$

where i is the load profile index

k is the load class index

n is the data index

The class k's average hourly load profile, represented by $TLP_k(t)$, can be calculated from the load profiles of the class using equation 15:

$$TLP_k(t) = \frac{\sum_{n=0}^{N} S_{k,i}(t)}{N} \qquad (15)$$

where k is the load class index,

i is the index for a load profile in load class k.

The proposed method can reconstruct a load profile $S'_{k,i}(t)$ to replace AMI data using equation 16:

$$S'_{k,i}(t) = \frac{TLP_k(t) \times P_{k,i}}{M_{k,i}} \qquad (16)$$

$$P_{k,i} = \max_{t \in [0,T]} S_{k,i}(t) \qquad (17)$$

$$M_{k,i} = \frac{\sum_{t=0}^{T} TLP_k(t) \times P_{k,i}}{\sum_{t=0}^{T} S_{k,i}(t)} \qquad (18)$$

where k is the load class index

i is the load profile index in load class k

T is the largest time point index

The method combines the individual load profile's annual peak load $P_{k,i}$ (equation 17), average load scale factor $M_{k,i}$ (equation 18) and the $TLP_k(t)$ (equation 15) of the class's normalized load profile.

This load model reduces the size of the original AMI load data even further as the model only stores two measurements, $P_{k,i}$ and $M_{k,i}$, for each customer and $k$ classified wavelet load models, $TLP_k(t)$. For the example study presented here using the 323 customers, the $k$ is set to 5. Table 3 shows the number of customers assigned to each of the 5 classified wavelet load models.

| k | Number of Customers |
|---|---|
| 1 | 89 |
| 2 | 166 |
| 3 | 39 |
| 4 | 14 |
| 5 | 15 |

**Table 3: Number of customers assigned to each classified wavelet load model class k**

For the classified wavelet load model, 234 (72% of 323) of the *SE* (equation 11) values are within 10% of the original AMI load total energy. The other 28% of the synthesized classified wavelet load profiles have 60% to 80% of the original load energy.

Table 4 shows several ranges of the classified wavelet load model's percentage error values, *PE(n)* (equation 12), and the percentage of classified wavelet load model data that is within each respective *PE(n)* range.

| PE(n) | % synthesized load profiles |
|---|---|
| 10% Less | 22% |
| 10% to 20% | 24% |
| 20% to 30% | 43% |
| over 30% | 11% |

**Table 4: Synthesized Load Profiles' *PE(n)***

As expected, the classified wavelet load model is not as accurate as the wavelet load model. The goal of this paper is to show that the AMI load profile's coarse-resolution representation can be used in load profile classification. It should be noted that the example given here has not been optimized relative to the number of classes.

IV. LOAD ESTIMATION WITH WAVELET BASED LOAD MODELS

In this section the three different load models (i.e., original AMI data, wavelet load model, and classified wavelet load model) for the 323 customers are integrated with a model of a real world circuit. The model of the circuit used in the study is shown in figure 13, in which the red dots represent the customer load points.

A time-series power flow analysis is performed for the 5028 hours from 5/27/2013 00:00 to 12/29/2013 13:00 for each of the three load models. The power flows at the start-of-the-circuit for phases A, B, and C are recorded for each hour, and the results of the wavelet models are compared with the results from the original AMI data.

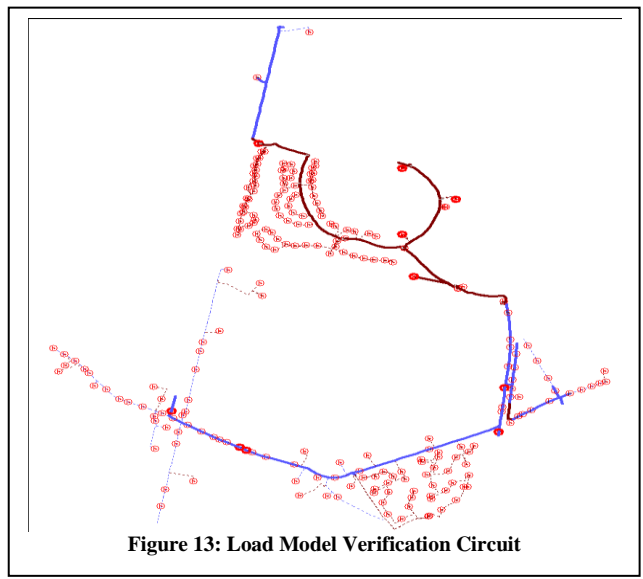

**Figure 13: Load Model Verification Circuit**

The wavelet load model stores 651 level 3 approximation coefficients for each of the 323 customers. The classified wavelet load model stores 2 values for each customer plus 5 different hourly TLPs (each has 5,028 data points). Table 5 presents the data compression that is achieved by each wavelet-based load model.

|  | No. of Wavelet Model Values | No. of AMI Values | % Compression |
|---|---|---|---|
| Wavelet Load Model | 210,273 | 1,682,184 | -87.50% |
| Wavelet Classified Load Model | 26,686 | 1,682,184 | -98.41% |

**Table 5: Data Compression Achieved by Wavelet Load Models**

Figure 14 shows a distribution of the hourly power flow percentage errors between the circuit with the wavelet load models and the circuit with the original AMI load data. From the figure it may be seen that 71% of the hourly power flows calculated with the wavelet load model are within 10% of the results using the original AMI data model. It may also be seen that 80% of the hourly power flows calculated with the classified wavelet load model are within 10% of the results using the original AMI data. Thus, using the original AMI data as a reference, the classified load model is providing more accurate results here than the wavelet model, which requires more storage.

Table 6 presents a comparison of circuit peak estimates obtained with each of the three load models. The original

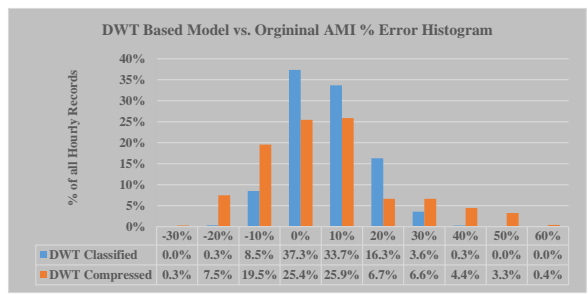

**Figure 14: Hourly Load Estimation % Error Histogram**

AMI data indicates the peak load time occurred on 7/18/2013 at 5PM for phases A and C phase, and on 7/18/2013 at 4PM for phase B. Both wavelet-based load model estimate the peak load occurred on 7/18/2013 at 5PM for all phases. Thus, there is a 1 hour difference in the phase B peak estimate. The peak load estimations from wavelet-base load models are within 12% of the peak estimate with the original AMI load data. Note that the wavelet load model provides a conservative estimate of the peak for all phases, whereas the classified wavelet load model underestimates the peak for all phases.

| Phase | AMI Peak (kW) | Peak Time | Wavelet Load Model Peak (kW) | Peak Time | Error % |
|---|---|---|---|---|---|
| A | 878.20 | 7/18 5 PM | 968.71 | 7/18 5 PM | 10.3% |
| B | 1132.82 | 7/18 4 PM | 1262.76 | 7/18 5 PM | 11.5% |
| C | 569.73 | 7/18 5 PM | 629.88 | 7/18 5 PM | 10.6% |
| Phase | AMI Peak (kW) | | Classified Wavelet Load Model Peak (kW) | | |
| A | 878.20 | 7/18 5 PM | 868.75 | 7/18 5 PM | -1.1% |
| B | 1132.82 | 7/18 4 PM | 1010.68 | 7/18 5 PM | -10.8% |
| C | 569.73 | 7/18 5 PM | 552.38 | 7/18 5 PM | -3.0% |

**Table 6: System Peak Load Estimation**

Table 7 shows that the total load estimates over the 5208 hours from the wavelet-based load models are within 1% of the estimates using the original AMI data.

| Phase | AMI (kWH) | Wavelet Load Model (kWH) | % Error |
|---|---|---|---|
| A | 2013192.71 | 2011098.14 | -0.10% |
| B | 2321937.76 | 2321775.12 | -0.01% |
| C | 1306861.05 | 1306961.94 | 0.01% |
| | | Classified Wavelet Load Model | |
| A | 2013192.71 | 2007990.47 | -0.26% |
| B | 2321937.76 | 2322710.13 | 0.03% |
| C | 1306861.05 | 1304125.65 | -0.21% |

**Table 7: Comparison of Wavelet Based Model Load Estimates for 5208 Hours with Estimate Using Original AMI Data**

## V. CONCLUSIONS

A major challenge to using big AMI data in large scale system power flow analysis is the size of the data. This paper evaluates two wavelet-based load models that can reduce the size of the AMI load data by approximately 88% (wavelet load model) and 98% (classified wavelet load model), respectively.

Individual customer load profiles can be synthesized by using lower resolution coefficients of the Discrete Wavelet Transformation. For the AMI loads considered here most of the energy is located in the lower frequency band, and with the DWT the energy of the lower frequency bands are captured by a set of approximation coefficients at level $k$, with a data size that is $2^{-k}$ of the original data size. The wavelet load model for k = 3 is investigated in this paper.

It has been shown that the wavelet representation of sudden/abnormal, sudden duration load changes can be detected in the higher frequency band modeled with the DWT detail coefficients. When the wavelet or classified wavelet load models are used, the large rapid load changes are filtered out.

A matrix derived from a 2-dimensional DWT has been used to model individual load profiles. The vertical coefficients of the matrix can be used to detect periodic load patterns, such as daily or weekly patterns, and thus to classify customer load profiles into groups that have similar load behavior. This provides a new approach to classifying customers into groups that display similar load behavior. Using the Typical Load Profile model of the class along with an individual customer's peak and average load measurements, a synthesized load profile can be reconstructed for each customer.

Wavelet-based load models have the potential to be a standard way to process and store AMI load data to be used in analysis. The wavelet load models investigated provide very accurate results, always within 0.26%, for system loss calculations. When estimating the peak, the wavelet models investigated are always within 11% of the AMI load data estimate, where the wavelet load model always provides a conservative estimate of the peak. Coupling the wavelet load models with actual SCADA measurements should overcome the peak estimate error and represents a next step in this investigation.

**Shiyin Zhong (M'08) received his B.S (2000) and MS (2003) degrees in computer engineering from Virginia Tech. His research interests include data mining, load forecasting and computer aided engineering.**

**Robert Broadwater is currently a faculty member in Virginia Tech. He received Ph.D. in electrical engineering from the Virginia Tech.**

**Steve Steffel (P.E). He is the manager of the distributed energy resources planning and analytics department in Pepco Holdings, Inc.**